\documentclass[aps,prb]{revtex4}

\usepackage{amsmath}
\usepackage{color}
\usepackage{amssymb}

\usepackage{graphicx}

\begin{document}

\title{Half - Quantum Vortices}

\author{V.P.Mineev}
\affiliation{Service de Physique Statistique, Magn\'{e}tisme et
Supraconductivit\'{e}, Institut Nanosciences et Cryog\'{e}nie, UMR-E CEA/UJF-Grenoble1, F-38054 Grenoble, France}

\begin{abstract}
Unlike to superfluid $^4$He the  superfluid $^3$He-A support the existence  of vortices with half quantum of circulation
as well as single quantum vortices. The singular single quanta vortices as well as nonsingular vortices with 2 quanta of circulation 
have been revealed in rotating $^3$He-A. However, the half quantum vortices in open geometry 
always possess an extra energy due to spin-orbit coupling leading to formation of domain wall at distances larger than dipole length $\sim 10^{-3}cm$ from the vortex axis. 
 Fortunately the same magnetic dipole-dipole interaction does not prevent the existence of half-quantum vortices in the polar phase of superfluid $^3$He  recently discovered  in  peculiar porous media 
"nematically ordered" aerogel. Here we discuss this exotic possibility. 

The discoveries of half-quantum vortices in triplet pairing superconductor Sr$_2$RuO$_4$ as well  in the exciton-polariton condensates are the other parts of the story about half-quantum vortices also described in the paper.

\end{abstract}

\date{\today}
\maketitle
\section{Introduction}
The quantization of circulation around vortex lines in superfluid $^4$He has been pointed out 
first by Lars Onsager in his famous remark \cite{Onsager} at the conference on statistical mechanics in Florence in 1949. The states of superfluid are described by the order parameter which is complex function $|\psi | e^{i\varphi}$, hence the phase $\varphi$ "may be multiple-valued, but its increment over any closed path must be a multiple of $2\pi$, so that the wave-function will be single-valued. Thus
the well known invariant called hydrodynamic circulation is quantized; the quantum of circulation is $h/m_4, . . .$" Indeed, the superfluid velocity is given by the gradient of  phase ${\bf v}_s=\hbar\nabla\varphi/m_4$, hence,  velocity circulation over a closed path $\gamma$ is
\begin{equation}
\Gamma=\oint_\gamma{\bf v}_sd{\bf r}=N\frac{h}{m_4}.
\end{equation} 
The quantized vortices differ each other by the integer number of circulation quanta N. In superfluid $^4$He the vortices with one quantum of circulation $h/m_4$ are usually created by the vessel rotation such that in equilibrium the total circulation around all vortices corresponds to the circulation of classic liquid rotating with given angular velocity.\cite{Feynman}  The energy of vortex per unit length 
\begin{equation}
E_v=\int \frac{\rho_sv_s^2}{2}d^2{\bf r}=\frac{\rho_s\Gamma^2}{4\pi}\ln\frac{R}{a}
\label{energy}
\end{equation}
is proportional to square of circulation, hence, the vortices with circulation quanta higher than 1 are unstable to decay for the vortices with one circulation quanta. Here, $R$ is  vessel size and $a$ is the coherence length which is of the order of the interatomic distance in liquid $^4$He. 

Magnetic field acts as rotation in case of charged superfluids, so in superconductors the quantized vortex lines also carry one quantum of magnetic flux $\phi_0=hc/2e$ and fixed value of magnetic moment $\phi_0/4\pi$.\cite{Abrikosov} The vortices with multiple flux quantum are energetically unstable in respect to decay to the single quantized vortices. This process can be written as sort of conservation law, for instance 2=1+1, of quanta of circulation.

The superfluid phases of $^3$He discovered in 1972 were proved to be much reacher  in respect of types of stable defects in the order parameter distribution.
For instance, in superfluid $^3$He-A there were predicted  4 type of stable vortices \cite{Volovik1976} with $N= 0, \pm 1/2, 1$ and the following algebra of 
addition of the circulations: 1+1=0, 1/2+1/2=1.  Half integer vortices in $^3$He-A till now were not registered. On the contrary,
half-integer flux quantization were observed in cuprate superconductors \cite{Kirtley} where it was proved to be
a powerful tool for probing the d-wave symmetry of the superconducting gap. 
The discovery of vortices with half-quantum flux has been reported recently in mesoscopic samples of spin-triplet superconductor 
Sr$_2$RuO$_4$ similar to superfluid $^3$He-A. \cite{Jang}  Even earlier half-quantum vortices have been revealed in quite different ordered media -  an exciton-polariton condensate.\cite{Lagourdakis}

Here we discuss the new possibility of half-quantum vortices realization that appeared
 with stabilization of polar phase of superfluid $^3$He in so called "nematically ordered" aerogel.\cite{Dmitriev12}.
 With this purpose in chapter 2 we introduce the notion of 
  half-quantum vortices in superfluid $^3$He-A. The advantages of realization of such type vortices in the superfluid polar state will be described in the following chapter. The story about discovery of half-quantum flux states in Sr$_2$RuO$_4$ is the subject of the 
 chapter 4. We conclude by mention of other even more exotic possibilities of half-vortices realization in a supersolid and in particular in Fulde-Ferrel-Larkin-Ovchinnikov  superconducting state.

\section{Vortices in superfluid $^3$He-A}

The matrix of order parameter of superfluid $^3$He-A 
\begin{equation}
A^A_{\alpha i}=\Delta V_\alpha(\Delta^\prime_i+i\Delta^{\prime\prime}_i)/\sqrt{2}
\label{op}
\end{equation}
is given \cite{Mineev} by the product of its spin and orbital parts. The unit spin vector ${\bf V}$ is situated in the plane perpendicular to the direction of spin up-up $|\uparrow\uparrow\rangle$ and down-down $|\downarrow\downarrow\rangle$ spins of the Cooper pairs. The vectorial product ${\bf \Delta}^\prime\times{\bf \Delta}^{\prime\prime}={\bf l}$
of orthogonal unit vectors  ${\bf \Delta}^\prime,{\bf \Delta}^{\prime\prime}$ determines the direction of the Cooper pairs orbital momentum  ${\bf l}$. The superfluid velocity in such a liquid is determined by
\begin{equation}
{\bf v}_s=\frac{\hbar}{2m_3}\Delta^\prime_i\nabla\Delta^{\prime\prime}_i.
\end{equation}
The velocity circulation is given by
\begin{equation}
\Gamma=N\frac{h}{2m_3}.
\end{equation}
The half-quantum vortices are admissible because
a change of sign of orbital part of the order parameter acquired  over  any closed path in the liquid 
corresponding to half quantum vortex can be compensated by the change of sign of the spin part of the order parameter, 
so that the whole order parameter will be single-valued. 
These vortices in the superflow field are simultaneously disclinations in the magnetic anisotropy field ${\bf V}$ with half -integer Frank index, analogous to the disclinations in nematic liquid crystals.

More visual picture of half-quantum vortices can be given assuming that all vectors   change their directions leaving in $(x,y)$ plane: $V_\alpha=\hat x_\alpha\cos\phi-\hat y_\alpha\sin\phi$, $\Delta^\prime_i+i \Delta^{\prime\prime}_i=(\hat x_i+i\hat y_i)e^{i\varphi}$.  Then the A-phase order parameter is written as $ A^A_{\alpha i}=\Psi^A_\alpha(\hat x_i+i\hat y_i)/\sqrt 2$, where
\begin{equation}
\Psi^A_\alpha=\Delta\left(e^{i\varphi_1}|\uparrow\uparrow\rangle_\alpha +e^{i\varphi_2}|\downarrow\downarrow\rangle_\alpha\right)/\sqrt{2},
\end{equation}
$|\uparrow\uparrow\rangle_\alpha=(\hat x_\alpha+i\hat y_\alpha)/\sqrt 2$,  $|\downarrow\downarrow\rangle_\alpha=(\hat x_\alpha-i\hat y_\alpha)/\sqrt 2$, $\varphi_1=\varphi+\phi,~\varphi_2=\varphi-\phi$. Thus, the order parameter of superfluid A-phase  is presented as the sum of the order parameters of spin up-up and spin down-down superfluids.  The single quantum vortex  corresponds to the order parameter distribution such that the phase increment of the orbital part of the order parameter  
over a closed path is $ \Delta\varphi=2\pi$. Here, the  both condensates with  up-up and  down-down spins acquires the same phase increment $\Delta \varphi_1=\Delta\varphi_2=2\pi$,
whereas the spin part of the order parameter is homogeneous ${\bf V}=const$. 
On the opposite, the half-quantum vortex is characterized by the  increments $\Delta\varphi=\pm\pi,~\Delta\phi=\pm\pi$. In two-condensates language this corresponds to the single quantum vortex either only in spin up-up $\Delta\varphi_1=\pm2\pi,~\Delta\varphi_2=0$, or only in spin down-down condensate that is $\Delta\varphi_2=\pm2\pi,~\Delta\varphi_1=0$. 

The gradient energy  in superfluid $^3$He is \cite{Mineev} 
\begin{equation}
{\cal F}_\nabla=\int d^3{\bf r}\left (K_1 \frac{\partial A_{\alpha i}}{\partial x_j}\frac{\partial A^\star_{\alpha i}}{\partial x_j}
+K_2 \frac{\partial A_{\alpha i}}{\partial x_j}\frac{\partial A^\star_{\alpha j}}{\partial x_i}
+K_3 \frac{\partial A_{\alpha i}}{\partial x_i}\frac{\partial A^\star_{\alpha j}}{\partial x_j}\right )
\end{equation}
It is easy to check that the energy corresponding to the combined defect consisting of half-quantum vortex in the orbital part of the order parameter  and the disclination in the vector $\bf V$ field is twicely smaller than the gradient energy of single quantum vortex.  More generally,
for superfliud phases with order parameter consisting of product orbital and spin vectors (\ref{op}) the energy of a defect is proportional to 
sum of squares of winding numbers of orbital and spin vector fields along a closed path around defect axis. For an half-quantum vortex it is $${\cal F}_\nabla=[(1/2)^2+(1/2)^2]\pi|\Delta|^2(2K_1+K_2+K_3)\ln\frac{R}{\xi}=\frac{\pi}{2}|\Delta|^2(2K_1+K_2+K_3)\ln\frac{R}{\xi},$$ whereas for a single quantum vortex it is  
$${\cal F}_\nabla=\pi|\Delta|^2(2K_1+K_2+K_3)\ln\frac{R}{\xi}.$$
 Thus, the half-quantum vortices looks as energetically more  profitable.

The singular single quanta vortices as well as nonsingular vortices with 2 quanta of circulation 
have been revealed in rotating $^3He-A$ (for review see \cite{Nature}) but half - quantum vortices were not. The reason for this is  the 
spin - orbital interaction caused by magnetic dipole interaction of Helium nucleus. In a superfluid phase with triplet pairing  the density of SO coupling energy is \cite{Mineev}
\begin{equation}
F_{so}=\frac{g_D}{5|\Delta|^2}\left ( A_{\alpha\alpha}A_{\beta\beta}^\star+A_{\alpha i}A^\star_{i\alpha}-\frac{2}{3}A_{\alpha i}A^\star_{\alpha i} \right ),
\label{so}
\end{equation}
that  in case of A-phase with order parameter (\ref{op})  is
\begin{equation}
F_{so}^A=\frac{g_D}{5}\left (
\frac{1}{3}-({\bf V}{\bf l})^2 \right ).
\end{equation}
Hence, at distances larger than dipole length $\sim 10^{-3}cm$ from the vortex axis the spin-orbital coupling suppress the inhomogeneity in the spin part of the order parameter distribution: vector $\bf V$ tends to be parallel or antiparallel to the direction of the Cooper pairs orbital momentum.  At these distances a disclination transforms in the domain wall (a planar soliton) possessing
energy proportional to its surface.\cite{Volovik1976,Volovik1977,Mineev98} The neutralization of the dipole energy can be reached in the parallel plate geometry where Helium fills the space between the parallel plates with distance smaller then dipole length under  magnetic field $H>>25$ G applied parallel to the normal to the plates.
This case the half quantum vortices can energetically compete with N=1 vortices. However, 
even in this case the  rotation of a "parallel plate" vessel with $^3$He-A will create lattice of half quantum vortices which at the same time presents  two-dimensional plasma of 
$\pm 1/2$ disclinations in the spin part of the order parameter with fulfilled condition of the "electroneutrality".\cite{Nature} The half quantum vortices in superfluid $^3He-A$ till now have not been revealed.

\section{Vortices in superfluid polar phase of $^3$He in "nematically ordered" aerogel}

Filling by liquid $^3$He an aerogel porous media allows to study influence of impurities on superfluid states with nontrivial pairing.\cite{Porto,Sprague}
There was found that both known in bulk liquid A and B superfluid phases of $^3$He also exist in aerogel.\cite{Barker} The new chapter in the investigations was opened when there was recognized that anisotropy of aerogel can influence superfluid $^3$He NMR properties. This way several states
of $^3$He-A with orbital and spin disordering have been discovered (see \cite{Dmitriev10} and references therein).
The following experimental investigations has been performed on $^3$He confined in a new type of aerogel consisting of Al$_2$O$_3$$\cdot$H$_2$O strands with a characteristic diameter $\sim 50$ nm and a chatacteristic separation of $\sim 200$ nm. The strands are oriented along nearly the same direction (say along $\hat z$ axis)   at macroscopic distance $\sim$ 3 - 5 mm that allows to call this aerogel as "nematically ordered" one. For liquid $^3$He in this type aerogel there were obtained indications that at low pressures the pure polar phase may exist in some range of temperatures just below critical temperature.\cite{Dmitriev12}

The pairing states of superfluid $^3$He in a random medium with global uniaxial anisotropy 
have been investigated by Aoyama and Ikeda \cite{Aoyama}. The corresponding second order  in the order parameter GL free energy density consists of isotropic part common for all the  superfluid phases with p-pairing and the anisotropic part
\begin{equation}
F^{(2)}=F^{(2)}_i+F^{(2)}_a=\alpha_0(T-T_c(x))A_{\alpha i}A^{\star}_{\alpha i}+\eta_{ij}A_{\alpha i}A^{\star}_{\alpha j},
\end{equation}
where the media  uniaxial anisotropy with anisotropy axis parallel to $\hat z$ direction is given by  the traceless tensor
 \begin{equation}
 \eta_{ij}=\eta\left (\begin{array}{ccc}1&0&0\\
 0&1&0\\
 0&0&-2\end{array}\right )
 \end{equation}
The B-phase  state  $A_{\alpha i}^B= \Delta R_{\alpha i} e^{i\varphi}$ is indifferent to the presence of uniaxial anisotropy $F^{(2)}_a(A_{\alpha i}^B)=0$, whereas the equal spin pairing states with the order parameter of the form $A_{\alpha i}=V_{\alpha}A_i$ creates the various possibilities.  

(i) A-phase $A_i=\frac{\Delta}{\sqrt{2}}(\hat x_i+i\hat y_i) $ 
\begin{equation}
F_a=\eta|\Delta|^2
\end{equation}

(ii)  A-phase $A_i=\frac{\Delta}{\sqrt{2}}\left (\hat z_i+i(\hat x_i\cos\alpha+\hat y_i\sin\alpha)\right ) $
\begin{equation}
F_a=-\eta|\Delta|^2/2
\end{equation}

(iii) Polar-phase $A_i=\Delta\hat z_i e^{i\varphi}$
\begin{equation}
F_a=-2\eta|\Delta|^2
\end{equation}

Which phase has the highest transition temperature from the normal state depends on the sign of $\eta$.   At negative $\eta<0$ the highest $T_c$ belongs to A-state (i)
with the Cooper angular momentum direction $\vec l$ parallel to the anisotropy axis.  At positive $\eta>0$ the preference has the polar state. Let us discuss now the mechanism to create the global anisotropy.

According to Rainer and Vourio \cite{Rainer}   the energy of a thin disk shape body immersed in $^3$He-A  depends of orientation of disk surface in respect to $\vec l$ vector and the minimum of this energy corresponds to the parallel orientation of the normal  to the disk surface to the $\vec l$. Hence if there are multiple disks homogeneously distributed in space with somehow fixed orientation parallel each other  this should stimulate the phase transition to the A-phase state with $\vec l$ parallel to the disks normal direction that corresponds to the $\eta<0$.

On the contrary  the most profitable orientation of a cigar shape object immersed in $^3$He-A is that  the cigar axis perpendicular to $\vec l$. The multiple cigars homogeneously distributed in space with axis parallel each other 
 should stimulate 
the A-phase state with $\vec l$ vectors randomly directed  in the plane perpendicular to the cigars axis  that corresponds to the $\eta>0$.

 The described difference between the order parameter orientations takes place so long we discuss only A-phase state.  The Rainer-Vuorio arguments 
extended to the other superfluid states show that the orientational energy of cigar tape objects immersed in the polar  phase can be even smaller than it is  for the A-phase with $\vec l$ perpendicular  to the cigars axis. Hence, the cigars type objects with parallel axis will stimulate phase transition to the polar state. 
It means that first there will be  phase transition to the polar state (iii) 
and then at lower temperature, when the fourth order terms in free energy are important,  one must expect the second order type phase transition to the distorted A-phase $A_i\propto \left (\hat z_i+ia(\hat x_i\cos\alpha+\hat y_i\sin\alpha)\right )$ 
transforming at low temperatures to the A-phase (ii) with vectors $\bf l$ randomly distributed in (x,y) plane as it was predicted by Aoyama and Ikeda. \cite{Aoyama}
  It is interesting that the similar phenomenon  with two subsequent phase transitions has been revealed  in 
multi-sublattice antiferromagnet  CsNiCl$_3$. \cite{Walker,Plumer}
As for superfluid $^3$He there were already obtained indications \cite{Dmitriev12} on existence of the polar state in "nematically ordered" aerogel.

Substituting the order parameter of polar state
\begin{equation}
A^{pol}_{\alpha i}=\Delta V_\alpha\hat z_ie^{i\varphi}
\end{equation}
in the expression (\ref{so}) for the spin-orbital energy density
we get
\begin{equation}
F_{so}^{pol}=\frac{2g_D}{5}\left (({\bf V}\hat z)^2-\frac{1}{3} \right ).
\end{equation}
We see that the spin-orbit coupling settles vector $\bf V$ in the plane perpendicular to the directions of aerogel strands. 
From this observation trivially follows that along with the singular vortices with phase $\varphi$ increment over any closed path
equal to a multiple of $2\pi$ there are half-quantum vortices with increment $\Delta\varphi=\pm\pi$ accompanied by disclination in the field $\bf V$ with Frank index 1/2.   According  the argumentation applied in previous chapter  to A-phase the half-quantum vortices in polar state are more energetically profitable than single quantum vortices. However, unlike A-phase where the spin-orbital coupling prevent existence of half-quantum vortices in rotating vessel 
this is not the case in the polar state. 

Thus,  rotation of vessel  filled by the superfluid polar phase of $^3$He in "nematically ordered" aerogel with angular velocity larger than the lower critical one must be accompanied by creation of half-quantum vortices as the most energetically profitable objects imitating rotation of the polar state superfluid component.

\section{Vortices in superconducting  strontium ruthenate} 

Sr$_2$RuO$_4$  is nonconventional superconductor possessing many unusual properties (for review see \cite{Maeno,Maeno2012}).
Common believe based on the absence of the Knight shift changes \cite{Ishida} below the critical temperature is that here we deal with superconductivity with triplet pairing. The material crystal structure is tetragonal with the point group symmetry $D_{4h}$. This case the order parameter for superconducting states with triplet pairing are related either to one-dimensional  representation or two dimensional representation of the point group.\cite{Samokhin}  For example, the order parameter for $A_{1u}$ representation is 
$A_{\alpha i}\hat k_i=|\Delta| \hat z_\alpha \hat k_ze^{i\varphi}$ and for $E_{u}$ representation is $A_{\alpha i}\hat k_i=|\Delta| \hat z_\alpha (\hat k_x+i\hat k_y)e^{i\varphi}$. In both cases the direction of spin vector ${\bf V}=\hat z$ fixed by spin-orbital coupling is pinned to the tetragonal axis.  If the spin part of the order parameter is fixed the only stable order parameter defects are the single flux quantum Abrikosov vortices. At the same time if one creates condition allowing  vector $\bf V$ free rotation like in superfluid phases of $^3$He one can expect the existence of half-quantum flux vortices. As we remember the energy of half-quantum vortices accompanied by disclination in the $\bf V$  field is smaller than the energy of single quantum vortex, but it is true only at the scale of distances from the vortex axis not exceeding the spin-orbital length.  At larger scales the increment  of spin-orbital energy due to vector $\bf V$  inhomogeneity will be larger than the  gain in gradient energy of half-quantum vortex in comparison with gradient energy of single quantum vortex with ${\bf V}|| \hat z$.
The spin-orbital length can be estimated in following manner.

The configuration ${\bf V}||\hat z$ means that the Cooper pair spins lie in the basal plane. Hence, below $T_c$  the magnetic susceptibility for the magnetic field oriented in basal plane should coincide with the susceptibility in the normal state and must decrease  for the field direction along the $c$ axis.\cite{Samokhin} In practice  it keeps the normal state value independently of field direction. There was found that the Knight shift is not changed for ${\bf H} \parallel \hat c$ for fields larger than 200G.\cite{Ishida} It means  this field  is already enough to rotate the Cooper pair spin system to be parallel or antiparallel to the field direction. In other words the 200 G field  is enough to overcome the the spin-orbital coupling. The comparison of corresponding paramagnetic energy with gradient energy of inhomogenious vector $\bf V$ distribution allows estimate the spin-orbital coherence length $\sim 50~\mu m$.  
So, to register the flux changes corresponding to half quantum vortices one must work with mesoscopic size samples. 

The authors of Science Report \cite{Jang} have used cantilever magnetometry to measure the magnetic moment of micrometer-sized  annual sample of strontium ruthenate prepared such that $ab$ crystal plane is parallel to  the plane of ring ($xy$ plane). The usual expression for the superfluid current density 
$$
\frac{4\pi\lambda^2}{c}{\bf j}=\frac{\phi_0}{2\pi}\nabla\varphi-{\bf A},
$$
 where $\lambda=\sqrt{mc^2/4\pi n_se^2}$ is the London penetration depth and $\phi_0=hc/2e$ is the flux quantum, leads to the fluxoid quantization which is the phase increment over a closed pass around the ring
 \begin{equation}
 \Phi^\prime=\frac{4\pi\lambda^2}{c}\oint{\bf j}d{\bf s}+\Phi=\phi_0N.
\end{equation}
Here $\Phi=\oint{\bf A}d{\bf s}$ is the magnetic flux. Then making use the expression for the ring magnetic moment 
$$\mbox{\boldmath$\mu$}=\int d^3{\bf r}({\bf r}\times{\bf j})/2c$$ one can write the ring magnetic moment for the magnetic field directed 
in $\hat z$ direction that is perpendicular to the ring plane 
\begin{equation}
\mu_z=\Delta\mu_zN+\chi_MH_z.
\end{equation}
Below the  lower critical field $H_{c1}=8~G$   the magnetic moment is the linear function of the external field with the negative slope
corresponding to the Meissner susceptibility $\chi_M$.  At each fields 8 G, 16 G, 24 G . . . as well at corresponding negative values of external field which are the multiples of the lower critical field there were revealed  the magnetic moment jumps  equal to $\Delta\mu_z=4.4\cdot 10^{-14}$ emu demonstrating penetration of single quantum vortices inside the ring. 

The crucial observation was obtained by application of field both in $\hat z$ and $\hat x$ directions. This case each jump in z-component of magnetic moment  starts to split  at increasing  $\hat x$ direction  field  component in two jumps  of twicely smaller heights $\Delta\mu_z=\frac{1}{2}\cdot 4.4\cdot 10^{-14}$ emu.
So, the experiment demonstrates the appearance of half flux quantum vortices. 

It is natural to ask why these vortices do not appear in the absence of $H_x$ field component. The plausible reason is that  the applied field in $\hat z$ direction does not exceed  50 G, which is probably smaller than necessary to overcome the spin-orbit coupling and settle vector $\bf V$ in the basal plane of crystal. On the contrary the 
measurements with $H_x$ field component were performed up to lower critical field of ring in $\hat x$ direction which is of the order of $\sim 250$ G. The jumps splitting  were distinguishable starting the fields $H_x\approx 80$ G that was enough to create  an inhomogeneous distribution  ${\bf V}=\hat z\cos\alpha+\hat y\sin\alpha$  with angle $\alpha$ increment equal to $\pm\pi$ along a closed path around the ring. 
The confirmation of the vector $\bf V$ nonhomogeneity  follows from calculation of  magnetic moment 
\begin{equation}
\mu_x=\frac{e}{2mc}\int d^3{\bf r}({\bf r}\times{\bf j}_s)_x/2c
\end{equation}
created by the spin current
\begin{equation}
{\bf j}_s=\hbar n_s\nabla\alpha.
\end{equation}
The estimation yields $\mu_x\approx 10^{-16}$ emu that corresponds to the measured value and points out that vector $\bf V$ is  indeed nonhomogeneously distributed  around the ring.

\section{Conclusion}

Each ordered media is characterized by particular type of coherence that can be probed through the  properties directly  
reflecting the symmetry and topology of ordering such as the  Josephson effect and quantized  vortices.
After discussion of  several instructive examples one can say that the situation when the order parameter of some ordered media consists of product of two parameters opens the possibility of existence of combined defects. Each part of such defect corresponds to the nonhomogeneous stable distribution of its  part of the total order parameter. In some particular cases like in polar state of superfluid $^3$He in "nematically ordered" aerogel or in mesoscopic superconducting rings of strontium rhuthenate these combined defects consist of half quantum vortex
and a disclination  with the Frank index 1/2 in the spin part of order parameter.

The half quantum vortices  have also  been observed in exciton-polariton condensate \cite{Lagourdakis}. The order parameter of this ordered media 
$${\bf  \Psi}={\bf e}_\lambda|\psi|e^{i\varphi}$$
is given by the product of condensate wave function and the vector of light polarization.
 Along with the ordinary vortices with phase increment along  a closed path a multiple of $2\pi$ the ordering like this obviously allows 
the combined defects  consisting of half quantum vortex and disclination with the Frank index 1/2 in the field of polarization vector 
${\bf e}_\lambda({\bf r})$.

Finally, it is worth to mention not yet discovered   half-quantum vortices in such ordered media as charge density waves
- CDW,
spin density waves - SDW, super solids, and Fulde-Ferrel-Larkin-Ovchinnikov - FFLO superconducting state.  For instance, in case of 2D periodic ordering in $(x,y)$ plane all of these "quantum crystal" orderings can be characterized by the order parameter of the form
$$\Psi=A\cos({\bf k}\mbox{\boldmath$\rho$}+\phi)e^{i\varphi}.$$
Then it is clear that the space increment of  each phase $\phi$ or $\varphi$ along a closed path can be multiple of $2\pi$ as well the multiple of $\pm\pi$. In the latter case the half quantum vortex in the field $\varphi({\bf r})$  should be accompanied by a half-quantum vortex
in the field $\phi({\bf r})$. More interesting possibilities one can find  in the paper
 by O. Dimitrova and M.V.Feigel'man \cite{Dimitrova}.

\end{document}